\documentclass[12pt]{article}
\usepackage{amssymb}
\usepackage[dvips]{graphicx}
\usepackage{epsfig}
\usepackage{epsfig}
\begin{document}
\setlength{\baselineskip}{0.75cm}
\setlength{\parskip}{0.45cm}
\begin{titlepage}
\begin{flushright}
DO-TH 97/09 \linebreak
May 1997
\end{flushright}
\vskip 1.2in
\begin{center}
{\Large\bf Leading log radiative corrections to deep inelastic 
production of heavy quarks}
\vskip 0.8in
{\large I.\ Schienbein}
\end{center}
\vskip .3in
\begin{center}
{\large Institut f\"{u}r Physik, Universit\"{a}t Dortmund \\
D-44221 Dortmund, Germany}
\end{center}
\vskip 1in
{\large{\underline{Abstract}}}

\noindent
${\cal O}(\alpha)$ QED radiative corrections to neutral current deep 
inelastic production of heavy quarks are calculated in the leading log 
approximation and compared with the corresponding corrections assuming a 
massless charm parton. Besides the inclusive case, corrections to 
the semi-inclusive
$d^3\sigma/dx dy dz$ and the effect of $z$-cuts are studied.
In the latter case, the massless corrections differ from
the correct massive radiative corrections to deep inelastic heavy 
quark production by about $40 \%$--$10 \%$ for 
$0.2\lesssim z \lesssim 0.5$. 
\end{titlepage}
\section{Introduction}
Deep inelastic production of heavy quarks at HERA is important for 
several reasons:
\begin{itemize}
\item At small $x$ ($x \approx 10^{-3}$) the charm contribution 
$F_2^c$ to $F_2^{ep}$ amounts to $20 \%$--$30 \%$ \cite{adloff96} 
making a reliable theoretical treatment of $F_2^c$ necessary for 
a precision measurement of $F_2^{ep}$.
\item The bulk of heavy quarks is produced via the photon gluon 
fusion mechanism \cite{adloff96,zeus97} providing the possibility to 
constrain the gluon distribution $g(y,\mu^2)$ in the proton 
\cite{vogt96,laenen96,daum96}.
\item By measuring differential distributions, the charm production 
mechanism can be tested, and it can be studied whether and when the 
charm quark behaves like a massless parton \cite{laenen96,daum96}.
\end{itemize}

In order to estimate radiative corrections to heavy quark production, 
we employ the well-known leading log approximation (LLA) 
\cite{km88,kms91,jbl90,jbl95}.
See also ref.~\cite{jbl91} where radiative corrections to heavy
flavour production have been studied in the Weisz\"acker-Williams
approximation (WWA) using hadronic variables.
The ${\cal O}(\alpha)$ (QED-)corrections to 
the process $e+g\longrightarrow e+c+\bar{c}$ are shown in 
fig.~\ref{Feynman} (not shown are crossed and virtual diagrams).
The main part of the corrections comes from bremsstrahlung from the 
electron line (plus corresponding virtual corrections), see a) and b). 
These two contributions are enhanced by a large collinear logarithm 
$L_e=\ln(Q^2/m_e^2)$. In the same way, calculating  c), d), and e), 
one gets logarithms $L_c=\ln(Q^2/m_c^2)$ which are much smaller than 
$L_e$ because of the large charm mass $m_c$, compared to the electron 
mass $m_e$. For example, for $Q^2=10$ GeV$^2$, one finds 
$L_e \approx 17.5>>L_c\approx 1.5$ ($m_c=1.5$ GeV).
Diagrams d) and e) do not contribute for another reason: the photons 
are not resolved from the final state jets in general.
(Only a light intermediary quark in diagram c) could give rise for a 
collinear singularity, which would have to be handled the same way as 
the corresponding collinear gluon emission, leading to a negligible 
(at least in LLA) modification of the parton distributions.)
Thus, we only take the leptonic corrections a) and b) 
(in ${\cal O}(\alpha)$-LLA) into consideration.

The paper is organized as follows: In Sect.~2 the formulae needed to 
calculate radiative corrections to neutral current heavy quark 
production in ${\cal O}(\alpha)$-LLA are collected. 
In Sect.~3 numerical results are presented.
Finally, the main results are briefly summarized in Sect.~4.
\section{Formalism}
We calculate the ${\cal O}(\alpha)$ QED-corrections to deep inelastic 
production of heavy quarks in leading log approximation (LLA) 
\cite{kms91, jbl90}.
Following the formalism of ref.~\cite{kms91}, the cross section for 
deep inelastic production of charm quarks via the photon gluon fusion 
(PGF) mechanism ($e+g\rightarrow e+c+\bar{c}$) reads:
\begin{equation}
d\sigma(ep\rightarrow ec\bar{c}X)=\int_0^1 dz_1\int_0^1dz_2
\int_0^1dz_3 D_{e/e}(z_1,Q^2)
g(z_2,\mu^2)\bar{D}_{e/e}(z_3,Q^2)
d\hat{\sigma}(e+g\rightarrow e+c+\bar{c})\ .
\label{wq}
\end{equation}
The functions $D_{e/e}$, $\bar{D}_{e/e}$ contain the factorized mass 
singularities from the electron line and are, up to ${\cal O}(\alpha)$, 
given by \cite{kms91}
\begin{eqnarray}
D_{e/e}(x,Q^2) = \bar{D}_{e/e}(x,Q^2)= \delta(1-x) + 
\left(\frac{\alpha L_e}{2\pi}\right)P_{ee}(x)
\label{dee}
\end{eqnarray}
with 
\begin{equation}
L_e\equiv \ln \frac{Q^2}{m_e^2}
\end{equation}
and the splitting function
\begin{equation}
P_{ee}(x) =  \left[\frac{1+x^2}{1-x}\right]_{+}= 
\frac{1+x^2}{(1-x)_{+}}+\frac{3}{2}\delta(1-x).
\end{equation}
The $(+)$-distribution is defined by
\begin{equation}
\int_x^1\ dz\ P_{ee}(z)\ f(z)=\int_x^1\ dz\ \frac{1+z^2}{1-z}\ 
(f(z)-f(1))-\int_0^x\ dz\ \frac{1+z^2}{1-z}\ f(1)\ .
\end{equation}
$D_{e/e}(z_1,Q^2)$ can be interpreted as the probability of finding an 
electron with momentum fraction $z_1$ 
inside the incoming electron, $\bar{D}_{e/e}(z_3,Q^2)$ describes the 
fragmentation of the scattered electron 
into the observed outgoing electron with momentum fraction $z_3$, and 
$g(z_2,\mu^2)$ is the gluon density inside the proton. Finally, 
$d\hat{\sigma}$ ist the cross section of the underlying hard scattering
subprocess.

The variables $x$, $y$ and $Q^2$ are reconstructed by the 4-momentum 
$p'_e$ of the observed electron
and the 4-momenta $p_e$, $p$ of the initial state electron and proton:
\begin{eqnarray}
q_{l}\equiv p_e-{p^{\prime}_e},\qquad Q_{l}^{2}\equiv -q_{l}^{2},
\qquad y_{l}
=\frac{p\cdot q_{l}}{p\cdot p_e},
\qquad x_{l}\equiv \frac{Q_{l}^{2}}{2p\cdot q_{l}}
=\frac{Q_l^2}{S y_l}\ ,
\label{outerkin}
\end{eqnarray}
where $S\equiv (p+p_e)^2$. In the same way, one 
defines for the hard scattering process
$e(\hat{p}_e=z_1 p_e)+g(\hat{p}=z_2 p)\longrightarrow e(\hat{p}^\prime_e
=p_{e}^{\prime}/z_3)+c(p_c)+X$: 
\begin{eqnarray}
\label{subkin}
\hat{q}&=&{\hat{p}_e}-{{\hat{p}^\prime_e}},\qquad 
\hat{Q}^{2}=-\hat{q}^{2}=\frac{z_1}{z_3}Q_l^2,\qquad
\hat{y}=\frac{\hat{p}\hat{q}}{\hat{p}{\hat{p}_e}}
=\frac{z_1z_3-1+y_l}{z_1z_3},
\\
\hat{x}&=&\frac{\hat{Q}^{2}}{2\hat{p}\cdot \hat{q}}
=\frac{\hat{Q}^2}{\hat{S} \hat{y}}
=\frac{z_1x_ly_l}{z_2(z_1z_3-1+y_l)},\qquad
\hat{S}=(\hat{p}_e +\hat{p})^2=z_1 z_2 S\ .
\end{eqnarray}
Because we only use leptonic variables $x_l$, $y_l$, etc.~, the index
$l$ will be suppressed from now on.

In the following, we calculate the radiative corrections to the photon 
gluon fusion (PGF) subprocess 
\begin{eqnarray}
e(\hat{p_e})+g(\hat{p})\longrightarrow e(\hat{p}_e^\prime)+c(p_c)
+\bar{c}(p_{\bar{c}}).
\label{pgf}
\end{eqnarray}
and compare the results with the radiative corrections to the charm 
excitation (CE) subprocess
\begin{eqnarray}
e(\hat{p_e})+c(\hat{p})\longrightarrow e(\hat{p}_e^\prime)+c(p_c) 
\label{eqeq}
\end{eqnarray}
where, in contrast to the PGF, the charm quark is treated as an
intrinsic (massless) parton of the proton.
The latter process has been used by the H1 Collab.~\cite{glazov} in a 
recent analysis of leptoproduction of D-mesons \cite{adloff96},
employing some charm density $c(x,\mu^2)$. 
\subsection{Inclusive case}
\subsubsection{Photon gluon fusion (PGF)}
The hard cross section for the process (\ref{pgf}) can be written as:
\begin{equation}
\frac{d^2\hat{\sigma}}{d\hat{x}d\hat{y}}=
\frac{4\pi\alpha^{2} \hat{S}}{{\hat{Q}}^4}
\left[(1-{\hat{y}})f_2\left({\hat{x}},{\hat{Q}}^{2}\right)
+{\hat{x}}{\hat{y}}^{2}f_1\left({\hat{x}},{\hat{Q}}^{2}\right)\right]
\label{partonwq}
\end{equation}
with the partonic structure functions
\cite{witten76,gr79,leveille79,lrsn93} 
\begin{eqnarray}
f_k({\hat{x}},{\hat{Q}}^2,m_c^2,\mu^2) & = & 
\frac{\alpha_s(\mu^2)}{\pi}\ \frac{\hat{\xi}}{4\pi}
\ e_c^2\ c_{k,g}^{(0)}({\hat{x}},{\hat{Q}}^2)\ 
\theta(\hat{W}^2-4m_c^2),\ \: k=2,L \: ,
\\
f_1({\hat{x}},{\hat{Q}}^2,m_c^2,\mu^2) & = & \frac{1}{2{\hat{x}}}
(f_2-f_L)\: ,
\\
\frac{\hat{\xi}}{4\pi}c_{2,g}^{(0)} & = & 
\left[\frac{{\hat{x}}}{2} -{\hat{x}}^2(1-{\hat{x}})+
2\frac{m_c^2}{{\hat{Q}}^2}{\hat{x}}^2(1-3{\hat{x}})
-4\frac{m_c^4}{{\hat{Q}}^4}{\hat{x}}^3\right]
\ln \frac{1+\hat{\beta}}{1-\hat{\beta}}\nonumber \\
& & +\hat{\beta}\left[4{\hat{x}}^2(1-{\hat{x}})
-\frac{{\hat{x}}}{2}-2\frac{m_c^2}{{\hat{Q}}^2}{\hat{x}}^2
(1-{\hat{x}})\right],
\\
\frac{\hat{\xi}}{4\pi}c_{L,g}^{(0)} & = & -4
\frac{m_c^2}{{\hat{Q}}^2}{\hat{x}}^3
\ln \frac{1+\hat{\beta}}{1-\hat{\beta}}+2\hat{\beta} {\hat{x}}^2
(1-{\hat{x}}), 
\label{HQProd}
\end{eqnarray}
where $e_c$, $m_c$ are the charm quark charge and mass, respectively, 
$\displaystyle \hat{\xi}\equiv \xi(\hat{Q}^2) = 
\frac{{\hat{Q}}^2}{m_c^2}$\,,\,$\displaystyle 
\hat{W}^2\equiv W^2(\hat{x},\hat{Q}^2)=\frac{\hat{Q}^2
(1-\hat{x})}{\hat{x}}$, and
$\hat{\beta}^2\equiv \beta^2(\hat{x},\hat{Q}^2)=1-4m_c^2/\hat{W}^2$.

By insertion of eq.~(\ref{dee}), (\ref{partonwq}) into eq.~(\ref{wq}),
one obtains the cross section 
$d\sigma=d\sigma ^0+d\sigma^{i}+ d\sigma^{f}$.
The Born cross section $d\sigma^0$ is given by
\begin{equation}
\frac{d^{2} \sigma^0}{dx dy}=\frac{4\pi\alpha^{2} S}{Q^4}
\left[(1-y)F_2\left(x,Q^{2}\right)+xy^{2}F_1\left(x,Q^{2}\right)\right]
\label{wqem}
\end{equation}
with the leading order structure functions 
\begin{eqnarray}
F_1(x,Q^2,\mu^2,m_c^2)=\int_{ax}^1\ \frac{dz_2}{z_2}\ g(z_2,\mu^2)
\ f_1(x/z_2,Q^2,m_c^2,\mu^2),
\\
F_2(x,Q^2,\mu^2,m_c^2)=\int_{ax}^1\ \frac{dz_2}{z_2} z_2\ g(z_2,\mu^2)
\ f_2(x/z_2,Q^2,m_c^2,\mu^2) ,
\end{eqnarray} 
where $a=1+4m_c^2/Q^2$.
$d\sigma^{i,f}$ contain the contribution due to radiation of a collinear
photon from the incoming (initial state radiation, ISR) and outgoing 
electron (final state radiation, FSR),
respectively:
\begin{eqnarray}
\frac{d^2\sigma^i}{dxdy}& = & \frac{\alpha L_{e}}{2\pi}
\int^1_{z_{1}^{min}}dz_1\left[\frac{1+z_1^2}{1-z_1}
(\sigma_0(z_1,1)-\sigma_0(1,1))\right] \nonumber
\\
&& +\ \frac{\alpha L_{e}}{2\pi} H(z_{1}^{min})\sigma_0(1,1),
\label{initial}
\end{eqnarray}
\begin{eqnarray}
\frac{d^2\sigma^f}{dxdy} & = & \frac{\alpha L_{e}}{2\pi}
\int^1_{z_{3}^{min}}dz_3\left[\frac{1+z_3^2}{1-z_3}
(\sigma_0(1,z_3)-\sigma_0(1,1))\right] \nonumber
\\
&& +\ \frac{\alpha L_{e}}{2\pi} H(z_{3}^{min})\sigma_0(1,1),
\label{final}
\end{eqnarray}
with
\begin{eqnarray}
\label{pgfs0}
\sigma_0(z_1,z_3) & = &\int^1_{z_{2}^{min}}dz_2\  g\left(z_2,\mu ^2\right)
\ \frac{\partial{(\hat{x},\hat{y})}}{\partial{(x,y)}}
\ \frac{d^2\hat{\sigma}}{d\hat{x}d\hat{y}}, 
\\
H(z) & = & -\int_0^z dx \frac{1+x^2}{1-x}=2\ln(1-z)+z+\frac{1}{2}z^2,
\end{eqnarray}
and the Jacobian
\begin{equation}
 \frac{\partial{(\hat{x},\hat{y})}}{\partial{(x,y)}}
=\frac{y}{z_2z_3(z_1z_3-1+y)}.
\label{jacobi1}
\end{equation}
The electromagnetic coupling is taken to be 
$\alpha\equiv \alpha(Q^2=m_e^2)=1/137.036$\footnote{
Using a running coupling $\alpha(Q^2)$ according to 
eq.~(14) in \protect\cite{kms91},  
taking effective quark masses \protect\cite{hector}
$m_u=m_d=0.041$, $m_s=.15$, $m_c=1.5$, and $m_b=4.5\ {\rm GeV}$, 
instead of a constant $\alpha$, leads to negligible differences.}.
The integration bounds
\begin{eqnarray}
\label{bounds}
z_{1}^{min} & = & \frac{1-y}{1-xy}+\frac{4m_c^2}{S(1-xy)},
\nonumber\\
z_{3}^{min} & = & \frac{1-y+xy}{1-4m_c^2/S},
\\
z_{2}^{min}(z_1,z_3)&=&\left(1+\frac{4m_c^2}{Q^2}\frac{z_3}{z_1}\right)
z_1xy\frac{1}{z_1z_3+y-1}
\nonumber
\end{eqnarray}
follow from the conditions $\hat{W}^2 \ge 4 m_c^2\ , 
\ 0 \le z_{2}^{min}(z_1,z_3) \le 1$. 
If no photon is radiated ($z_1=1$, $z_3=1$), one finds the well-known
expression $z_{2}^{min}(1,1)=(1+4m_c^2/Q^2)x\equiv a x$.
In the kinematical region of small $x$, the bounds read approximately: 
$z_{1}^{min}= z_{3}^{min}= 1-y+{\cal O}(x)+{\cal O}(m_c^2/S)$.
It should be remarked that $\sigma_0$ is related to the Born 
cross section by $d^2\sigma^0/dxdy=\sigma_0(1,1)$.
\subsubsection{Charm excitation (CE)}
In the case of electron quark scattering, the radiative corrections 
in LLA are known up to
${\cal O}(\alpha^2)$ \cite{km88,kms91,jbl90,jbl95,hector}.
For comparison with eq.~(\ref{initial}), (\ref{final}), we use 
equations (22)--(29) in \cite{kms91},
with the charm quark as the only massless parton (MP) in the 
initial state. 
Furthermore, contributions due to $Z$-exchange can be neglected 
in the realm of $Q^2 \le 50$ GeV$^2$, i.~e.~, we make the
replacements $A^c\to e_c^2$, $B^c\to 0$ 
(in eq.~(22)--(29) in \cite{kms91}).
For the partonic structure functions, the simple relations
\begin{eqnarray}
f_2(\hat{x},\hat{Q}^2) & = & e_c^2\ \delta(1-\hat{x})\ ,
 \\
f_1(\hat{x},\hat{Q}^2) & = & \frac{1}{2\hat{x}}\ f_2(\hat{x},\hat{Q}^2)
\label{mpstrukt}
\end{eqnarray}
hold, and one easily finds \cite{kms91}
\begin{equation}
\sigma_0(z_1,z_3)=\frac{2\pi\alpha^2 y}{z_1 z_3^2 \hat{y}^3 \hat{S}}
[1+(1-\hat{y})^2]
e_c^2\left(c(\bar{z_2},\hat{Q}^2)+\bar{c}(\bar{z_2},\hat{Q}^2)\right)
\label{mps0}
\end{equation}
with
\begin{equation}
\bar{z}_2=\frac{z_1 x y}{z_1 z_3 + y -1}\ , \ z_{1}^{min}
=\frac{1-y}{1-xy}\ , \ z_{3}^{min}=1-y(1-x)\ .
\label{mpbounds}
\end{equation}
(Of course, eq.~(\ref{mps0}) and (\ref{mpbounds}) have to be 
inserted into eq.~(\ref{initial}) and (\ref{final}).)

Finally, we briefly discuss the Compton contribution $d\sigma^C$ 
to the radiative corrections
which can be obtained from eq.~(21) ($Q_0=200\ {\rm MeV}$) 
and (29) in \cite{kms91}, only 
allowing for the charm (or anti-charm) quark in the initial state.
Fig.~\ref{compton} displays the Compton 
contribution $\displaystyle \delta^C=\frac{d^2\sigma^C}{dxdy}
/\frac{d^2\sigma^0}{dxdy}$ for $x=10^{-2}$ (dashed line) and $x=10^{-3}$
(full line) using the CTEQ4L parton distributions \cite{cteq4}.
For $y\lesssim 0.7$, $\delta^C$ is small and can be safely neglected for 
experimentally relevant values of $y$ \cite{adloff96}.
\subsection{$z$-differential case}
In order to calculate radiative corrections to the $z$-differential 
cross section 
$\displaystyle d^3\sigma^D/dx dy dz$, where $z=p\cdot p_D/p\cdot q$, 
the fragmentation of the charm quark into 
the observed $D$-meson must be taken into consideration.
Thus, it is necessary to extend eq.~(\ref{wq}):
\begin{equation}
d\sigma(ep\rightarrow eDX)=\int_0^1 dz_1\int_0^1dz_2
\int_0^1dz_3 D_{e/e}(z_1,Q^2)
g(z_2,\mu^2)\bar{D}_{e/e}(z_3,Q^2)
\int_0^1dz_4 D_c(z_4)d\hat{\sigma} .
\label{fragds}
\end{equation}
The hadronization of the outgoing charm quark with momentum $p_c$ 
into the observed $D$-meson 
with momentum $p_D=z_4p_c$ is modeled by the fragmentation 
function $D_c(z_4)$.
Writing the partonic cross section differential in 
$\hat{z}_c=\hat{p}\cdot p_c/\hat{p}\cdot \hat{q}$, the relations in
eq.~(\ref{subkin}) have to be accomplished with
\begin{equation}
\hat{z}_c=\frac{z y z_3}{z_4(z_1z_3+y-1)}\equiv \frac{z}{z_4}r,
\label{subkin2}
\end{equation}
with $r$ defined by 
\begin{equation}
r\equiv \frac{y\ z_3}{z_1z_3+y-1},
\label{r}
\end{equation}
which can be derived from the definitions of $\hat{z}_c$, $z$ 
and $\hat{q}$, using $p_D=z_4 p_c$.

In analogy to the above discussion,
we compare the ``massive'' (PGF) with the ``massless'' (CE) 
corrections, using a massless charm parton (MP).
\subsubsection{Photon gluon fusion}
The partonic cross section reads
\begin{equation}
\frac{d^3\hat{\sigma}}{d\hat{x}d\hat{y}d\hat{z}_c}=
\frac{4\pi\alpha^{2} \hat{S}}{{\hat{Q}}^4}
\left[(1-{\hat{y}})f_2\left({\hat{x}},{\hat{Q}}^{2};\hat{z}_c\right)
+{\hat{x}}{\hat{y}}^{2}f_1\left({\hat{x}},{\hat{Q}}^{2};
\hat{z}_c\right)\right],
\end{equation}
where the structure functions are given by
\footnote{The corresponding expressions for
general boson gluon fusion can be found in \protect\cite{ks97}.}
\begin{eqnarray}
f_2(x,Q^2;{\zeta}) &=& \frac{\alpha_s}{\pi}e_{c}^2x
\left[\frac{1}{4}{\rm BG}\left(x,Q^2,{\zeta},{m_c}\right)
+\frac{3}{4}{\rm BL}\left(x,Q^2,{\zeta},{m_c}\right)\right],
\label{dpgfdz1}\\
f_L(x,Q^2;{\zeta})&=&\frac{\alpha_s}{\pi}e_{c}^2x\frac{1}{2}
{\rm BL}\left(x,Q^2,{\zeta},{m_c}\right),
\label{dpgfdz2}
\end{eqnarray}
with
\begin{eqnarray}
{\rm BL}\left(x,Q^2,{\zeta},{m_c}\right)&=&
4x\left[1-x-x\frac{{m_c}^2}{Q^2}\frac{1}{{\zeta}(1-{\zeta})}\right],
\nonumber\\
{\rm BG}\left(x,Q^2,{\zeta},{m_c}\right)&=&-2+
\left[1-2x+2x^2+4\frac{{m_c}^2}{Q^2}x(1-x)\right]
\frac{1}{{\zeta}(1-{\zeta})}
\nonumber\\
&&+2x^2\frac{{m_c}^2}{Q^2} \left(1-2\frac{{m_c}^2}{Q^2}\right)
\frac{1}{(1-{\zeta})^2{\zeta}^2},
\nonumber
\end{eqnarray}
\cite{lrsn93,schuler88} and
$\displaystyle f_1(\hat{x},\hat{Q}^2;\hat{z}_c)= \frac{1}{2\hat{x}}
\left[f_2(\hat{x},\hat{Q}^2;\hat{z}_c)-
f_L(\hat{x},\hat{Q}^2;\hat{z}_c)\right]$.

The cross sections for ISR and FSR have the same structure
as in eq.~(\ref{initial}) and (\ref{final}) with a $z$ dependent 
function $\sigma_0$:
\begin{eqnarray}
\frac{d^3\sigma^i}{dxdydz} & = & \frac{\alpha L_{e}}{2\pi}
\int^1_{z_{1}^{min}}dz_1\left[\frac{1+z_1^2}{1-z_1}
(\sigma_0(z_1,1;z)-\sigma_0(1,1;z))\right] \nonumber
\\
&& +\ \frac{\alpha L_{e}}{2\pi}H(z_{1}^{min})\sigma_0(1,1;z),
\label{fragini}\\
\frac{d^3\sigma^f}{dxdydz} & = & \frac{\alpha L_{e}}{2\pi}
\int^1_{z_{3}^{min}}dz_3\left[\frac{1+z_3^2}{1-z_3}
(\sigma_0(1,z_3;z)-\sigma_0(1,1;z))\right] \nonumber
\\
&& +\frac{\alpha L_{e}}{2\pi}H(z_{3}^{min})\sigma_0(1,1;z),
\label{fragfin}
\end{eqnarray}
with
\begin{eqnarray}
\sigma_0(z_1,z_3;z) & = &\int^1_{z_{2}^{min}}dz_2 g
\left(z_2,\mu ^2\right)\int_{z_{4}^{min}}^{z_{4}^{max}}dz_4
\frac{\partial{(\hat{x},\hat{y},\hat{z_c})}}{\partial{(x,y,z)}}
\frac{d^3\hat{\sigma}}{d\hat{x}d\hat{y}d\hat{z_c}}D_c(z_4)\ .
\label{frags0}
\end{eqnarray}
The Jacobian in eq.~(\ref{frags0}) can easily be calculated as
\begin{eqnarray}
\frac{\partial{(\hat{x},\hat{y},\hat{z_c})}}{\partial{(x,y,z)}}=
\frac{\partial{(\hat{x},\hat{y})}}{\partial{(x,y)}}
\frac{\partial \hat{z}_c}{\partial z}=
\frac{y^2}{z_2z_4(z_1z_3-1+y)^2}.
\end{eqnarray}
The integration bounds can be derived from the conditions
$\displaystyle \hat{W}^2=\frac{\hat{Q}^2(1-\hat{x})}{\hat{x}} 
\ge 4m_c^2$, leading to eq.~(\ref{bounds}), and 
$\displaystyle \hat{z}_{c}^{min}\le \hat{z}_c \le 
\hat{z}_{c}^{max}$ with
$\displaystyle \hat{z}_{c}^{max \atop min}=
\frac{1\pm \beta(\hat{x},\hat{Q}^2)}{2}$ \cite{schuler88}, where
$\beta^2(\hat{x},\hat{Q}^2)=1-4m_c^2/{\hat{W}}^2$, implying
\begin{eqnarray}
z_{4}^{max}&=&\min \left[1,\frac{z r}{\hat{z}_{c}^{min}}\right],
\\
z_{4}^{min}&=&\min \left[z_{4}^{max},\frac{z r}{\hat{z}_{c}^{max}}
\right]
\end{eqnarray}
with $r$ from eq.~(\ref{r}).
For $D_c(z)$, we use a Peterson et al.\ fragmentation function 
\cite{peterson83}
\begin{equation}
D_c(z) = N \left\{ z \left[ 1-z^{-1}-\varepsilon_c/(1-z)
\right]^2\right\}^{-1}\ ,
\label{peterson}
\end{equation}
normalized to $\int_0^1 dz D_c(z) = 1$, i.~e.~,
\begin{equation}
N^{-1} =
\frac{({\varepsilon_c}^2-6{\varepsilon_c}+4)}{(4-{\varepsilon_c}) 
\sqrt{4{\varepsilon_c}-{\varepsilon_c}^2}}
\left\{
\arctan\frac{{\varepsilon_c}}{\sqrt{4{\varepsilon_c}
-{\varepsilon_c}^2}}
+ \arctan\frac{2-{\varepsilon_c}}{\sqrt{4{\varepsilon_c}
-{\varepsilon_c}^2}} \right\}
+ \frac{1}{2} \ln {\varepsilon_c} + \frac{1}{4-{\varepsilon_c}}\ .
\end{equation} 

\subsubsection{Flavor excitation}
In the subprocess $e+c\longrightarrow e+c$ we have
$\displaystyle \hat{z}_c=\hat{p}\cdot p_c/\hat{p}\cdot \hat{q}=1$, 
following from $p_c=\hat{q}+\hat{p}$.
Thus, one obtains the $\hat{z}_c$-differential cross section 
by inserting the structure functions
\begin{eqnarray}
f_2(\hat{x},\hat{Q}^2,\hat{z}_c) & = & 
e_c^2\ \delta(1-\hat{x})\delta(1-\hat{z}_c)
 \\
f_1(\hat{x},\hat{Q}^2,\hat{z}_c) & = & 
\frac{1}{2\hat{x}}\ f_2(\hat{x},\hat{Q}^2,\hat{z}_c)
\label{mpstruktdz}
\end{eqnarray}
into eq.~(\ref{partonwq}).
Because of $\displaystyle \delta(1-\hat{z}_c)=z_4\delta(z_4-r z)$ and
$\displaystyle \partial\hat{z}_c/\partial z=r/z_4$ one finds
\begin{equation}
\sigma_0(z_1,z_3;z)=\sigma_0(z_1,z_3)\ r D_c(rz)
\label{mps0dz}
\end{equation}
with $\sigma_0(z_1,z_3)$ from eq.~(\ref{mps0}) and $r$ 
from eq.~(\ref{r}).

Finally the $z$-differential corrections are given by 
eq.~(\ref{fragini}) and (\ref{fragfin}) 
with $\sigma_0$ from eq.~(\ref{mps0dz}) and the integration bounds
\begin{eqnarray}
z_{1}^{min}&=&\max\left[\frac{1-y}{1-xy},1-y(1-z)\right],
\\
z_{3}^{min}&=&\max\left[1-y(1-x),\frac{1-y}{1-yz}\right] .
\label{mpboundsdz}
\end{eqnarray}
The second boundary in $\max[ \ldots , \ldots ]$ can be deduced 
from $0 \le rz \le 1$.

Eq.~(\ref{fragini}), (\ref{fragfin}) can easily be tested:
\begin{equation}
\int_0^1 dz\frac{d^3\sigma^{i,f}}{dxdydz}\stackrel{!}{=}
\frac{d^2\sigma^{i,f}}{dxdy}\ ,
\end{equation}
with $d^2\sigma^{i,f}/dxdy$ from (\ref{initial}), (\ref{final}).
The Compton contribution will be neglected for the same reasons as 
in the inclusive case.
\section{Numerical results}
As usual, the radiative corrections will be shown in form of a 
correction factor $\delta$ defined by 
$d\sigma=d\sigma^0(1+\delta)$, i.~e., 
$\delta=\delta^i+\delta^f+\ldots$
with 
\begin{displaymath}
\delta^{i,f}(x,y)=\frac{d^2 \sigma^{i,f}}{dx dy}/
\frac{d^2 \sigma^0}{dx dy}
\end{displaymath}
or, in the $z$-differential case, 
\begin{displaymath}
\delta^{i,f}(x,y,z)=\frac{d^3 \sigma^{i,f}}{dx dy dz}/
\frac{d^3 \sigma^0}{dx dy dz}.
\end{displaymath}
In all figures we use HERA centre-of-mass energies 
$S=4\cdot27.5\cdot 820$ GeV$^2$.

Fig.~\ref{rcpgf} shows the radiative corrections to heavy quark 
production (PGF) for experimentally relevant values\footnote{For 
a clean extraction of the 
gluon density it is necessary to extend the $x$-range 
to $x\lesssim 5 \cdot 10^{-4}$ \cite{daum96}.} 
of Bjorken-$x$ \cite{adloff96} as a function of $y$ according to
eq.~(\ref{initial}) and (\ref{final}) using 
the GRV94(LO) parton distributions \protect\cite{grv94}.
The factorization scale has been chosen to be 
$\mu^2=Q^2+4m_c^2$ \cite{vogt96} 
with $m_c=1.5\ {\rm GeV}$.
For $x= 10^{-2}$, the typical shape of radiative corrections 
in leptonic variables can be seen, with large corrections 
for $y\to 1$, whereas for smaller $x\le 10^{-3}$ the curves 
become more and more flat for $y \to 1$.

The theoretical uncertainties due to different choices of 
parton distributions, factorization scales, and charm masses turn 
out to be small, as can be seen in fig.~\ref{pdfcomp} 
and fig.~\ref{compare}.
One finds $\delta({\rm CTEQ})-\delta({\rm GRV})< 0.02$,
$\delta(m_c=1.3)-\delta(m_c=1.7)< 0.03$ for relevant $y\le 0.7$, and
$\delta(\mu^2=4m_c^2)-\delta(\mu^2=Q^2+4m_c^2)< 0.03$, where the 
scale $\mu^2=4m_c^2$ has been favored in \cite{grs94}.
This could be expected because variations of $\mu$, $m_c$, or 
the parton distributions lead to rather similar changes in 
$d\sigma^0$ and $d\sigma^{i,f}$, so that the quotient $\delta$
does not change too much.

In the recent analysis of deep inelastic charm production by the 
H1 Collab.~\cite{adloff96}, the radiative corrections have been 
calculated in ${\cal O}(\alpha)$-LLA with the 
HECTOR package \cite{hector} using the charm excitation 
subprocess ($F_L^c=0$) and the GRV92 parton distributions \cite{grv92}.
Furthermore, only initial state radiation has been taken into 
consideration, assuming the collinear final state photon not to be 
separated from the outgoing electron \cite{glazov},
leading to small corrections. Because PGF has been measured to be 
the dominant 
($>95 \%$) charm production mechanism in the small
$x$ (and $Q^2$) range \cite{adloff96}, it is necessary to check 
if the ``massless'' corrections ($\delta^{MP}$) agree with the 
``massive'' ones ($\delta^{PGF}$).
In fig.~\ref{mpcpgfgrvi} we compare the massive 
($\mu^2=Q^2+4m_c^2$; $m_c=1.5$ GeV) 
with the massless corrections due to initial state radiation.
The experimentally relevant values of $Q^2$ \cite{adloff96} are
indicated by dotted vertical lines. 
The conversion of ``masslessly corrected'' ($\delta^{MP}$) data to 
``massively corrected'' ($\delta^{PGF}$) data
can be performed by applying the factor 
$R\equiv (1+\delta^{MP})/(1+\delta^{PGF})$ because of
\begin{eqnarray*}
d\sigma({\rm PGF})&=&d\sigma({\rm exp})\frac{1}{1+\delta^{\rm PGF}}
\\
&=&d\sigma({\rm MP})\frac{1+\delta^{\rm MP}}{1+\delta^{\rm PGF}}
\equiv d\sigma({\rm MP})R\ .
\end{eqnarray*}
For the $(x,Q^2)$ data points one finds $|R-1|=|\delta^{MP}-
\delta^{PGF}|/(1+\delta^{PGF})< 3\%$,i.~e.~, the differences 
between massless and massive radiative corrections lead to a small
increase ($\delta^{MP}>\delta^{PGF}$) of the data.
However, this difference is small enough
to use the simpler charm excitation subprocess for calculating 
the radiative corrections in the inclusive case.
%

Of course, heavy quark production processes are exclusive in
the heavy quark momentum and on this more differential level the
photon gluon fusion and the charm excitation processes are not 
compatible which can be seen, e.g., from the different shapes
of the $x_D=|\vec{p}_D^*|/|\vec{p}_p^{\, *}|=2|\vec{p}_D^{\, *}|/W$ 
distributions  shown in the experimental analyses
\cite{adloff96,zeus97} (fig.~6, fig.~1 resp.).
We prefer to employ the lorentz-invariant variable
$z=p\cdot p_D/p\cdot q$ which approximately transforms
into $x_D$ in the $\gamma^{\, *}$-$p$ centre-of-mass 
system, following the argumentation in \cite{adloff96}.
This can be easily seen by calculating $z$
in the $\gamma^{\, *}$-$p$-CMS. One finds $z=x_D \sin^2 \theta^{\, *}/2$
with\footnote{$\vec{p}_{\gamma}^{\, *}+\vec{p}_{p}^{\, *}=0$, 
$\vec{p}_{g,c}^{\, *}=z_2\vec{p}_{p}^{\, *} \Rightarrow 
\vec{p}_{\gamma}^{\, *}+\vec{p}_{g,c}^{\, *}=-(1-z_2)\vec{p}_{p}^{\, *}$.
This means for the
CE-subprocess: $\vec{p}_{\gamma}^{\, *}+\vec{p}_{c}^{\, *}=\vec{p'}_{c}^{\, *}
=z_4 \vec{p}_D^{\, *} \Rightarrow \theta^{\, *}=\pi$
and for the PGF: $\vec{p}_{\gamma}^{\, *}+\vec{p}_{g}^{\, *}=\vec{p}_{c}^{\, *}
+\vec{p}_{\bar{c}}^{\, *}
\Rightarrow \theta^{\, *}\approx \pi$ 
(the collinear configuration is dominant). Note that an 
angle $\theta^{\, *}=160^{\circ}$ still yields 
$\sin^2 \theta^{\, *}/2=0.97$.} 
 $\theta^{\, *}=\angle (\vec{p}_p^{\, *},\vec{p}_D^{\, *})
\approx \pi$. 

In fig.~\ref{mpcpgffrag} we show the  $z$-differential radiative 
corrections according to eq.~(\ref{fragini}), (\ref{fragfin}), 
(\ref{frags0}) (PGF, full line) and
(\ref{mps0dz}) (MP, dashed line), where it has been integrated 
over the kinematical range 
$10\ {\rm GeV}^2\le Q^2\le 100\ {\rm GeV}^2,\ 0.01 \le y\le 0.7$.
The left hand side shows the correction factor 
$\delta^{i}=d\sigma^{i}/d\sigma^0$ as well as
the sum of initial and final state radiation $\delta^i+\delta^f$. 
In all cases, we employed the GRV92(LO) (massless) 
\protect\cite{grv92} and GRV94(LO) (massive) \protect\cite{grv94} 
parton distributions, in the massive case
with $m_c=1.5$ GeV and $\mu^2=Q^2+4m_c^2$. For $D_c$ 
a Peterson et al.\ fragmentation function \protect\cite{peterson83} 
with $\varepsilon=0.15$ has been taken.
To study the dependence of $\delta$ on $\varepsilon$, we 
compare the radiative corrections for two
different choices of $\varepsilon$ in fig.~\ref{epscomp}.
In the massive case the $\varepsilon$-dependence is obviously small, 
whereas for massless corrections
one finds big differences in the steep region $z\lesssim 0.4$.
Fig.~\ref{mpcpgffrag} reveals a rather big difference between 
massive and massless corrections for $z\lesssim 0.5$.
For $z\to 1$ only soft photon radiation is allowed, so that the 
corrections factorize and become independent of the 
underlying subprocess.
On this more differential level it is necessary to
calculate the radiative corrections using the photon gluon fusion 
subprocess because of the big deviations of the massless
from the massive corrections for $z\lesssim 0.5$.
On the right side of fig.~\ref{mpcpgffrag} the $z$-dependence of the 
cross sections $d\sigma^{0,i,f}$ is displayed.
Integration over $0\le z\le 1$ leads to the observed small 
differences between massless and massive corrections because 
positive and negative contributions compensate each 
other. To get an impression of the effect of $z$-cuts, 
we show in fig.~\ref{zcut1} 
the correction factor $\delta^{i,f}(x,y;z^{min},z^{max})=
\int_{z^{min}}^{z^{max}}dz \frac{d\sigma^{i,f}}{dx dy dz}/
\int_{z^{min}}^{z^{max}}dz \frac{d\sigma^{0}}{dx dy dz}$ 
for three $z$-integration ranges, from top to bottom 
$z^{min}=0 \le z\le z^{max}=1$, $0.2 \le z\le 0.8$, and
$0.3 \le z \le 0.9$ for $x=10^{-3}$.
For completeness, we used the GRV94/92(LO) parton distributions 
\cite{grv94}, \cite{grv92} for the massive ($m_c=1.5$ GeV, 
$\mu^2=Q^2+4m_c^2$) and massless corrections
and a Peterson et al.\ fragmentation function \cite{peterson83} with 
$\varepsilon=0.15$.
As can be seen from fig.~\ref{mpcpgffrag}, a cut $z\gtrsim 0.2$
leads to the exclusion of positive contributions of $d\sigma^{i,f}$, 
diminishing the radiative corrections.
\section{Summary}
The ${\cal O}(\alpha)$-QED corrections to inclusive and 
$z$-differential deep inelastic electroproduction of heavy
quarks have been calculated in the leading log approximation,
using electron variables.
The results have been compared to the radiative corrections in the 
MP scheme, where the charm quark is assumed to be a massless parton
in the proton.
In the inclusive case, the differences between these two approaches 
turned out to be negligible.
However, the measurement of heavy quark production is of course
differential in the momentum of the (observed) heavy quark, 
recommending to perform the radiative corrections on the same
differential level.
Thus, we have considered the semi-inclusive $z$-differential 
case, in which the massive 
corrections have to be applied, i.e., using the photon gluon fusion 
subprocess, because for $0.2 \lesssim z \lesssim  0.5$ the 
massless corrections differ from the massive ones by about
$\approx 40 \%$--$10 \%$.
Furthermore, we studied the effect of cuts on the $z$-integration range. 
A cut $z \gtrsim 0.2$, e.g., 
excludes positive contributions of $d\sigma^{i,f}$, so that
the ($z^{min}\le z\le z^{max}$)-integrated corrections are smaller as 
in the fully inclusive, i.e., ($0\le z\le 1$)-integrated case.
\section*{Acknowledgements}
We thank E.\ Reya and M.\ Gl\"{u}ck for advice and useful discussions.
%

\newpage
%
%
%
%
%
\begin{figure}[h]
\vspace*{1cm}
  \centering
  \begin{tabular}{cc}
    \includegraphics[ bb= 50 50 550 500,width=4.5cm]{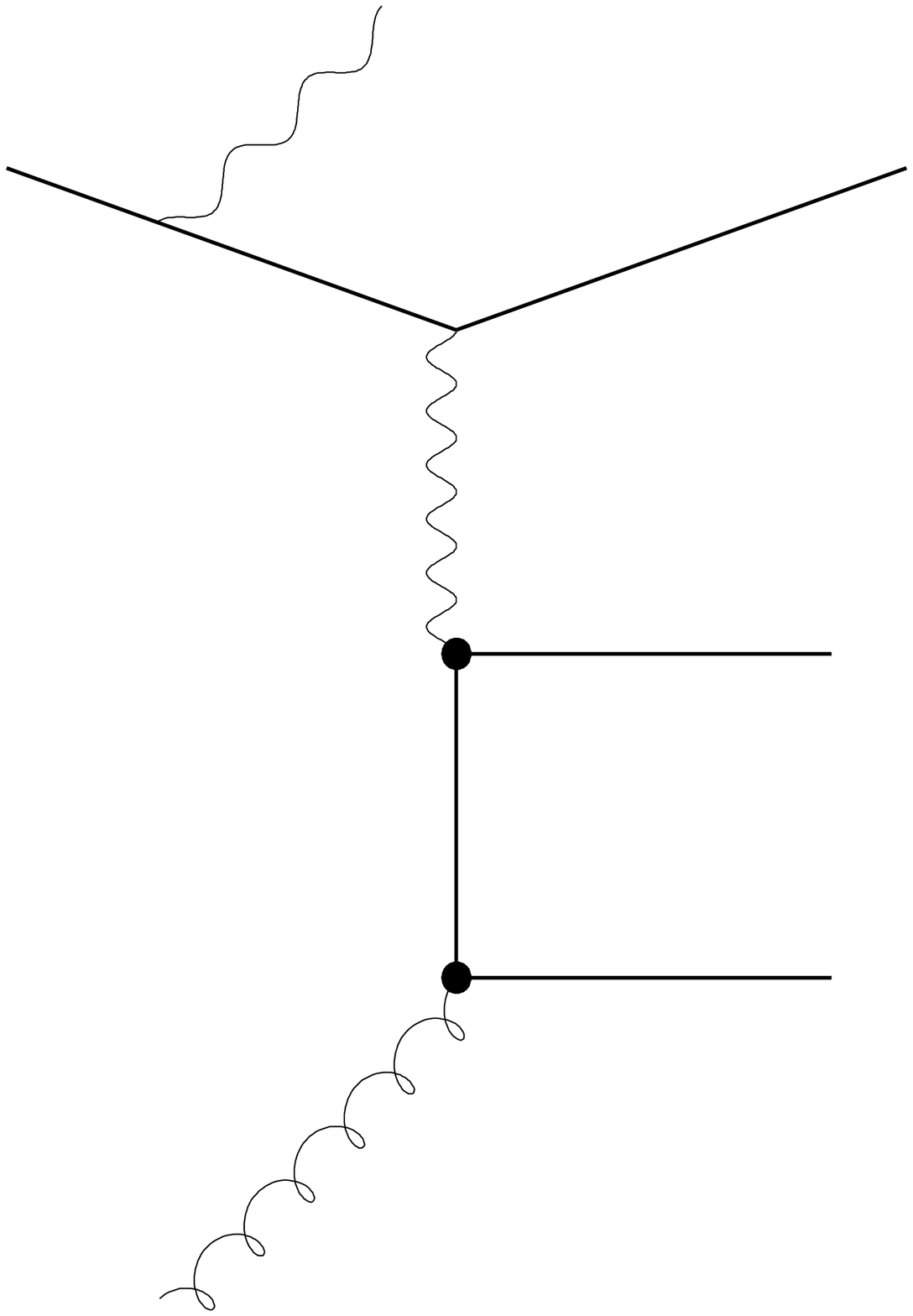}&
    \includegraphics[ bb= 50 50 550 500,width=4.5cm]{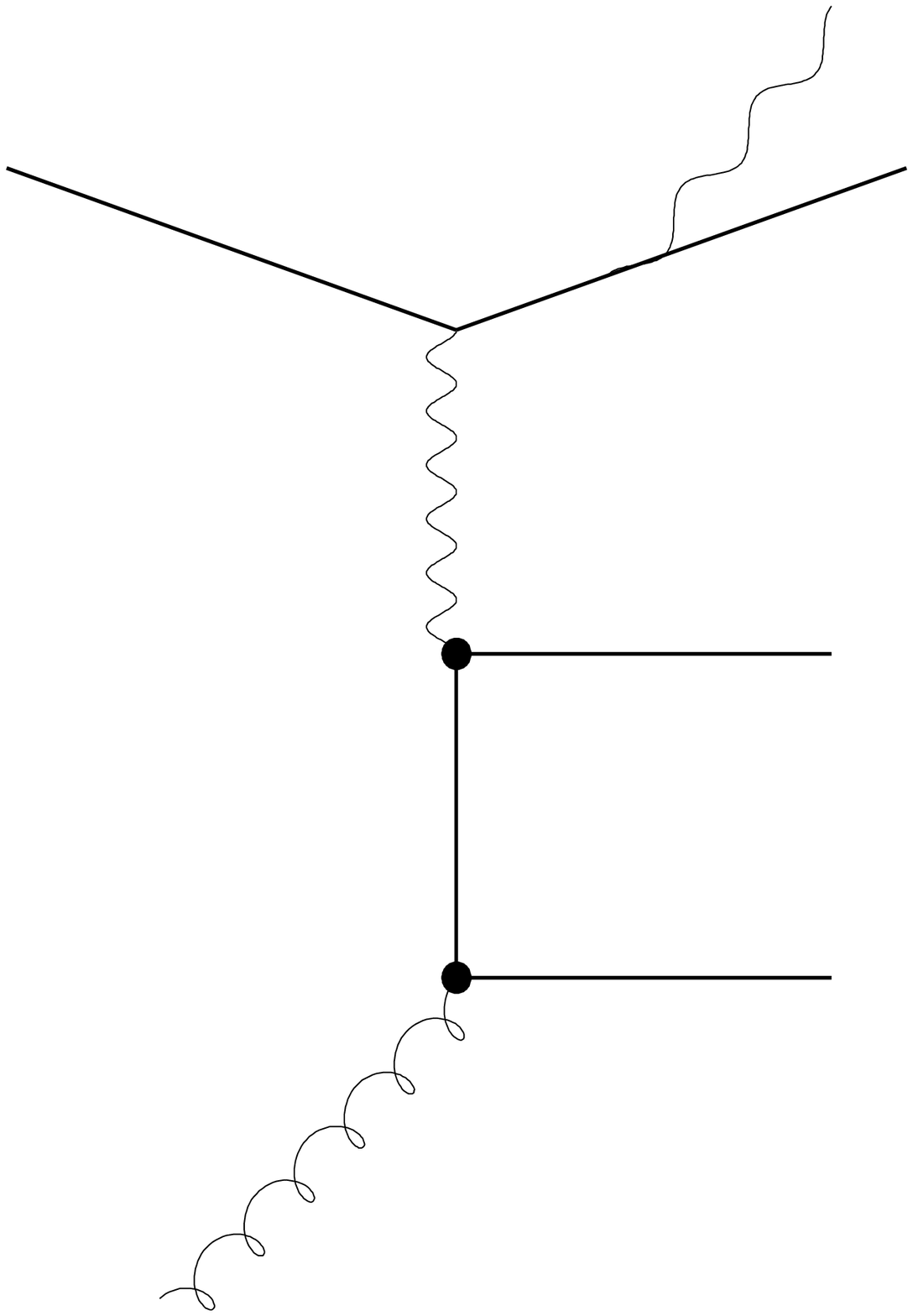}
\cr
    \large
    a)&\large b)\cr
    \normalsize
  \end{tabular}
\end{figure}

\begin{figure}[h]
  \centering
  \begin{tabular}{ccc}
    \includegraphics[ bb= 50 50 550 500,width=4.5cm]{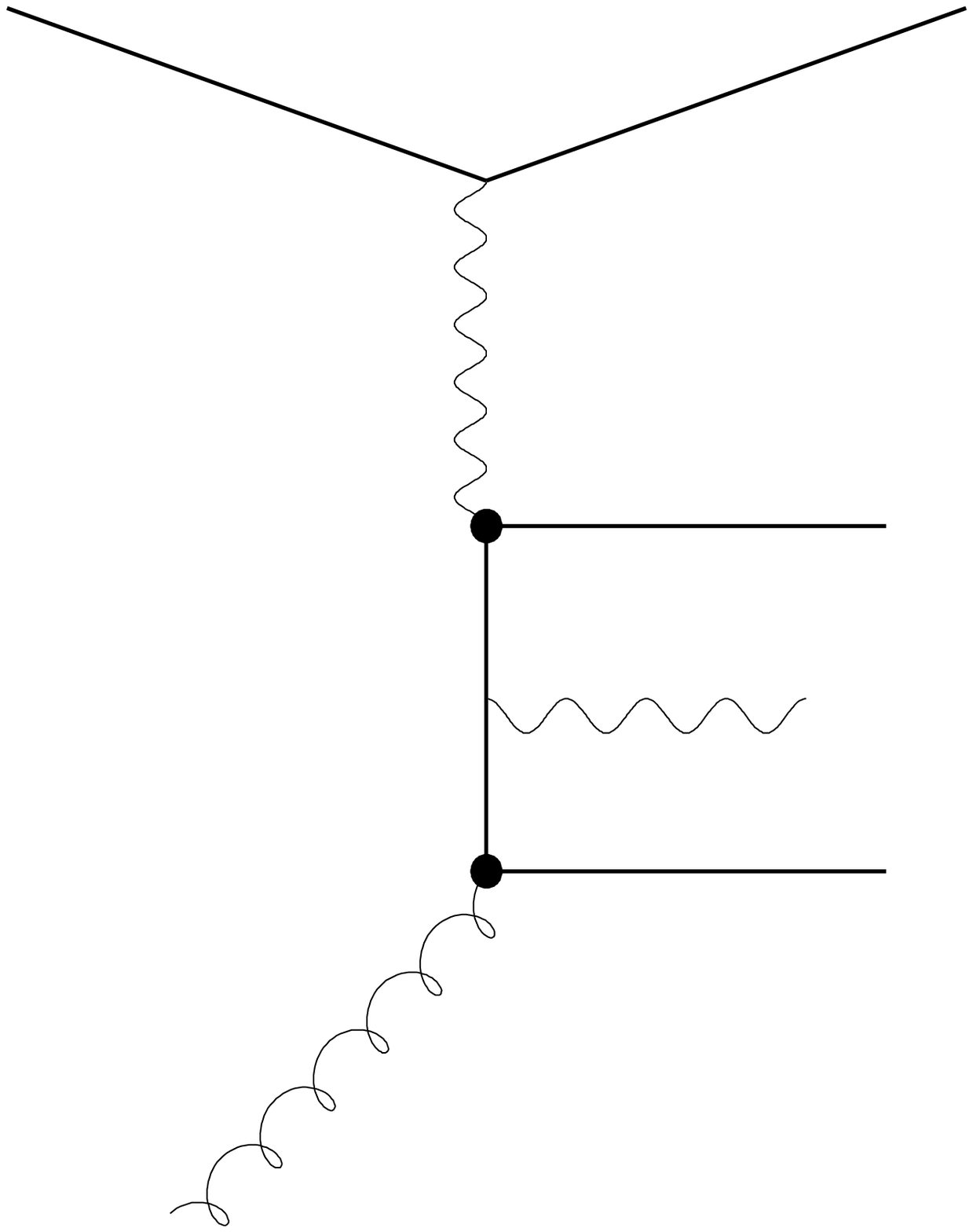}&
    \includegraphics[ bb= 50 50 550 500,width=4.5cm]{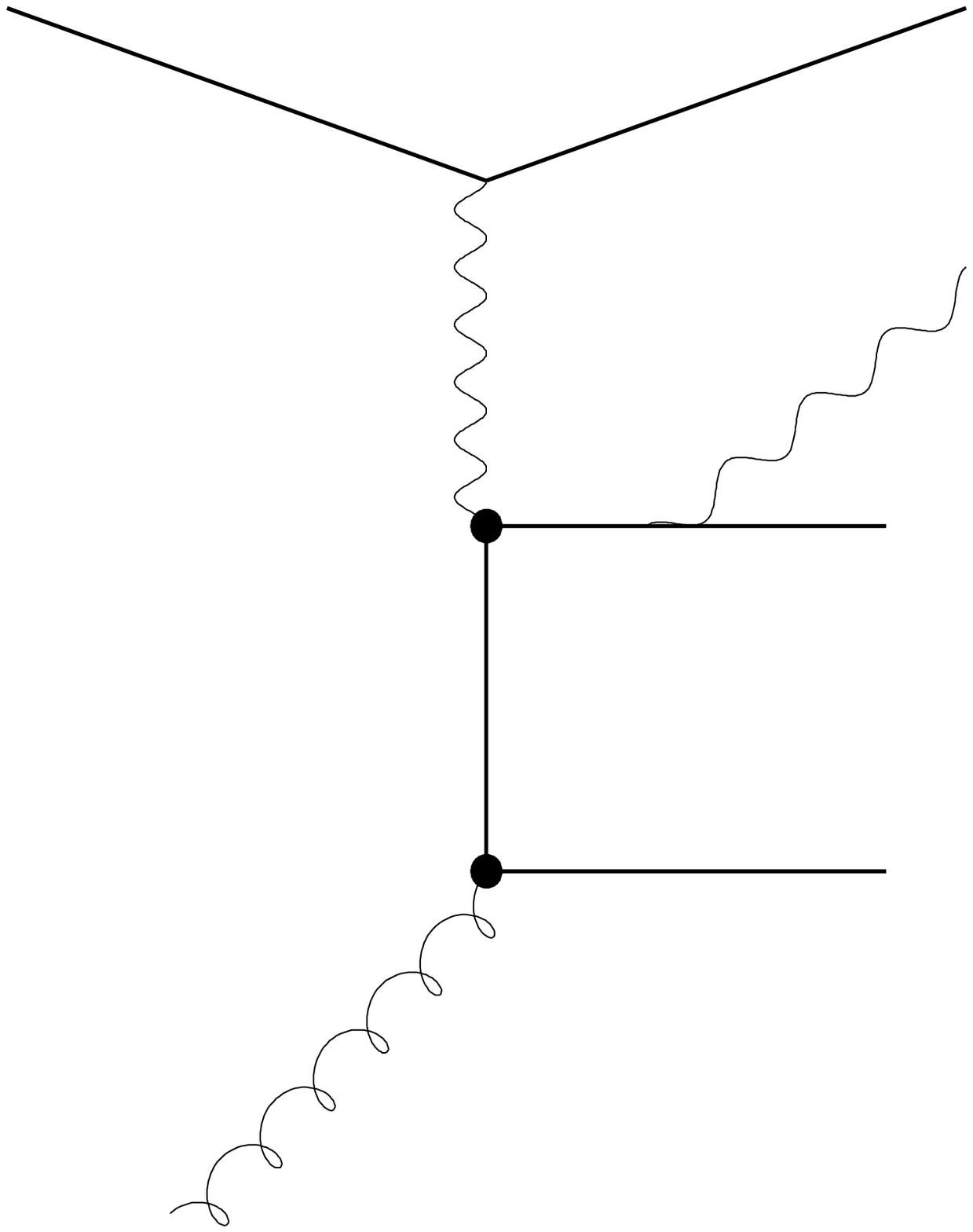}&
    \includegraphics[ bb= 50 50 550 500,width=4.5cm]{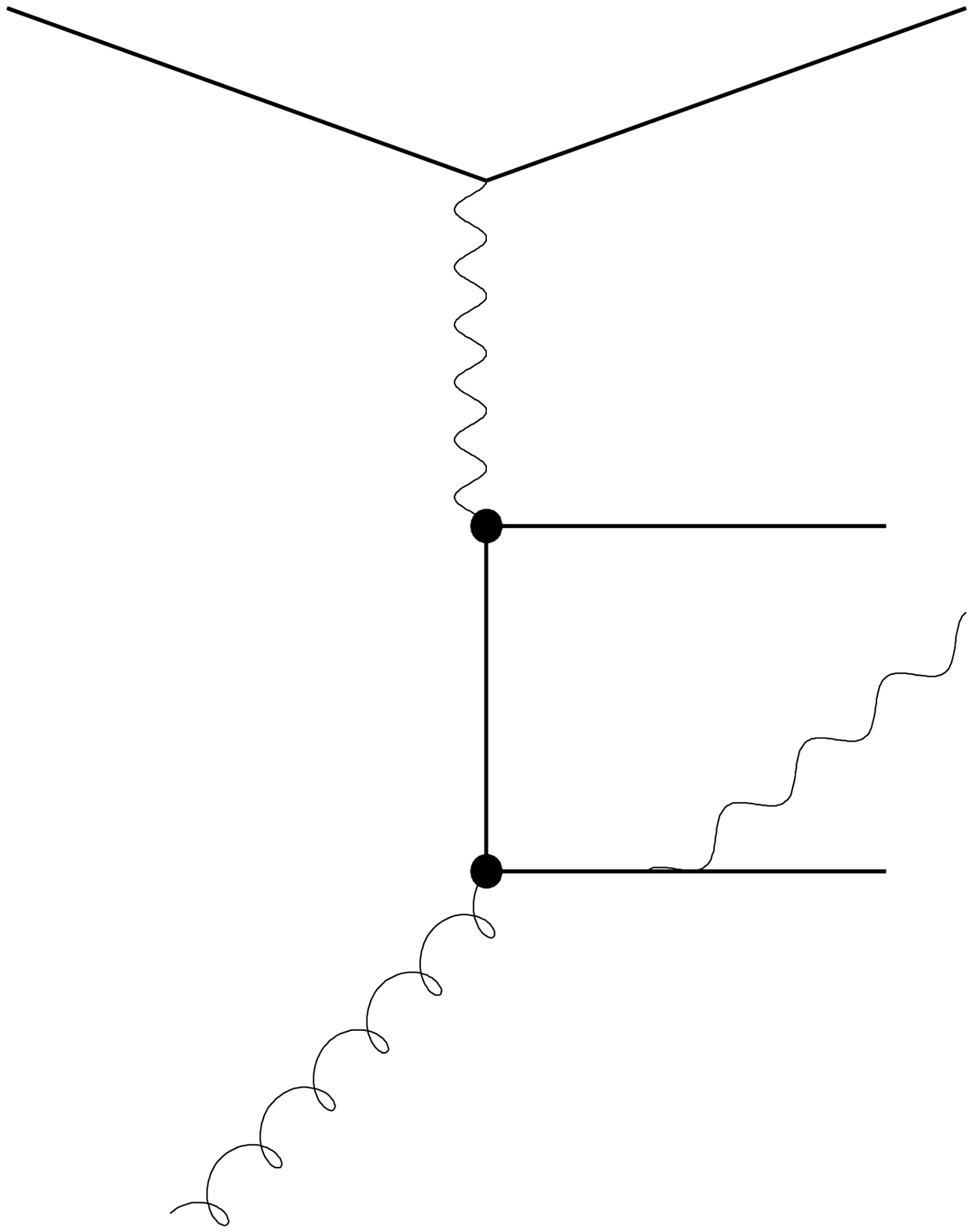}
\cr
    \large
    c)&\large d)&\large e)\cr
    \normalsize
  \end{tabular}
\caption{\sf Real ${\cal O}(\alpha)$ corrections to deep inelastic 
production of heavy quarks via the process 
$e+g\longrightarrow e+c+\bar{c}$. Not shown are the corresponding 
crossed diagrams.}
\label{Feynman}
\end{figure}

\begin{figure}[t]
\centering
\epsfig{figure=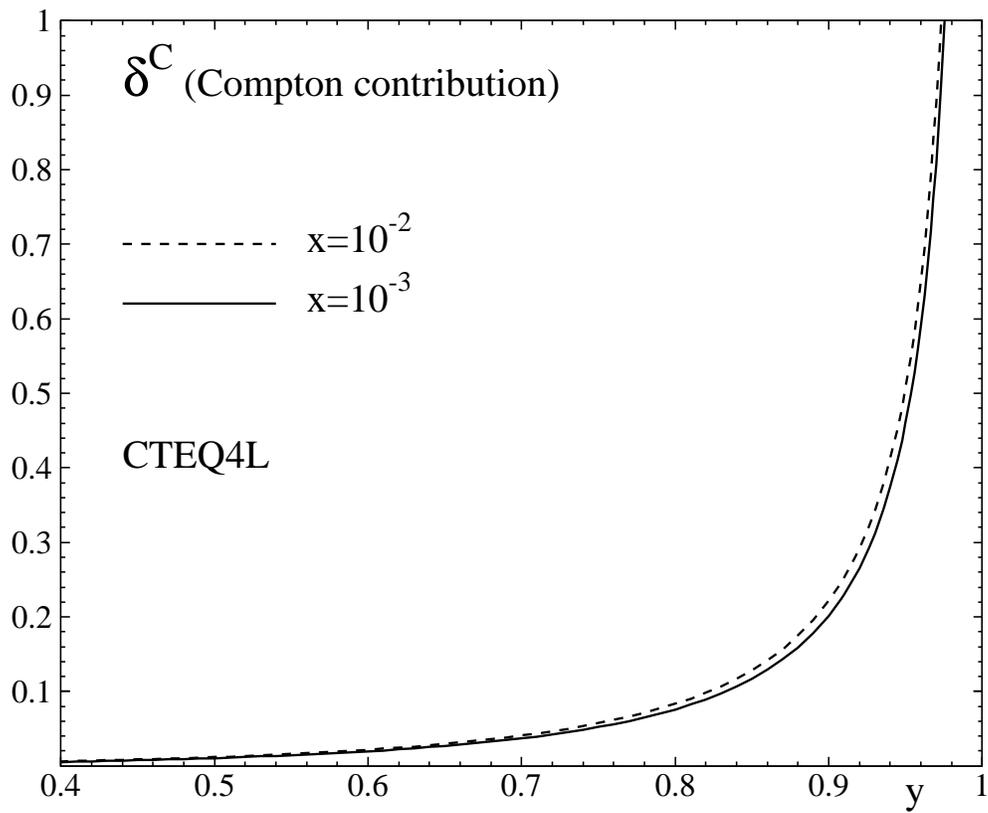,width=14cm}
\caption{\sf Compton contribution $\delta^{C}$ according to eq.~(29)
in \protect\cite{kms91} ($Q_0=200\ {\rm MeV}$, considering only 
the (anti-)charm quark in the initial state) with $x=10^{-3}$ 
(full line) and $x=10^{-2}$ (dashed line), employing the
CTEQ4L parton densities \protect\cite{cteq4}.}
\label{compton}
\end{figure}

\begin{figure}[ht]
\centering
\epsfig{figure=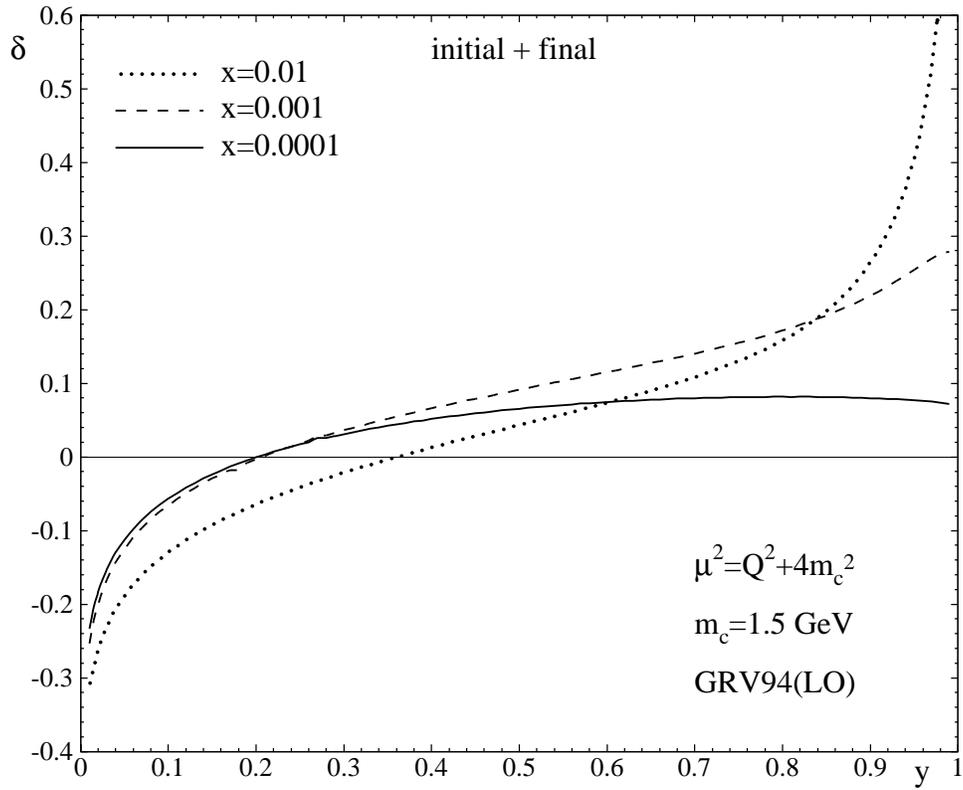,width=14cm}
\caption{\sf
Radiative corrections to heavy quark production (PGF) in 
${\cal O}(\alpha)$-LLA, using the GRV94(LO) parton distributions 
\protect\cite{grv94} and the factorization scale
$\mu^2=Q^2+4m_c^2$ with $m_c=1.5\ {\rm GeV}$.
(''initial'' and ``final'' refers to initial and final 
state radiation.)}
\label{rcpgf}
\end{figure}

\begin{figure}[ht]
\centering
\epsfig{figure=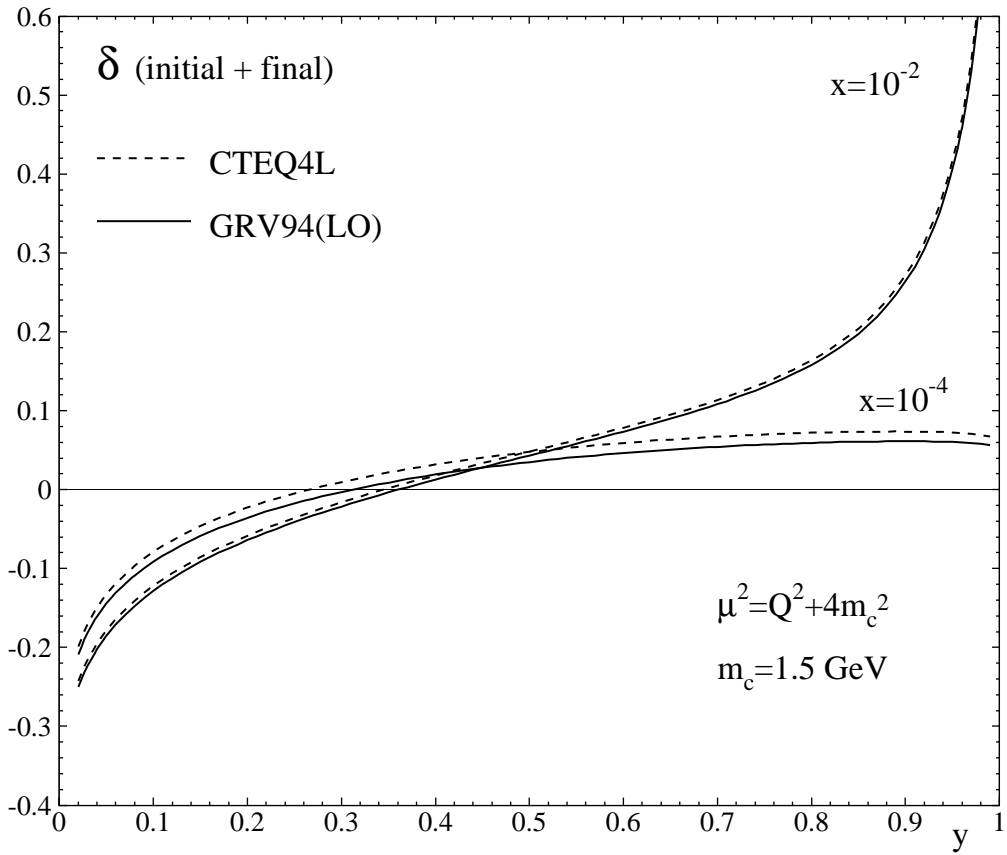,width=14cm}
\caption{\sf $\delta(x,y)$, using the GRV94(LO) \protect\cite{grv94} 
(full line) and CTEQ4L parton densities \protect\cite{cteq4} 
(dashed line) with $x=10^{-2}$ and $x=10^{-4}$, $\mu^2=Q^2+4m_c^2$ 
and $m_c=1.5\ {\rm GeV}$.}
\label{pdfcomp}
\end{figure}

\begin{figure}[ht]
\centering
\epsfig{figure=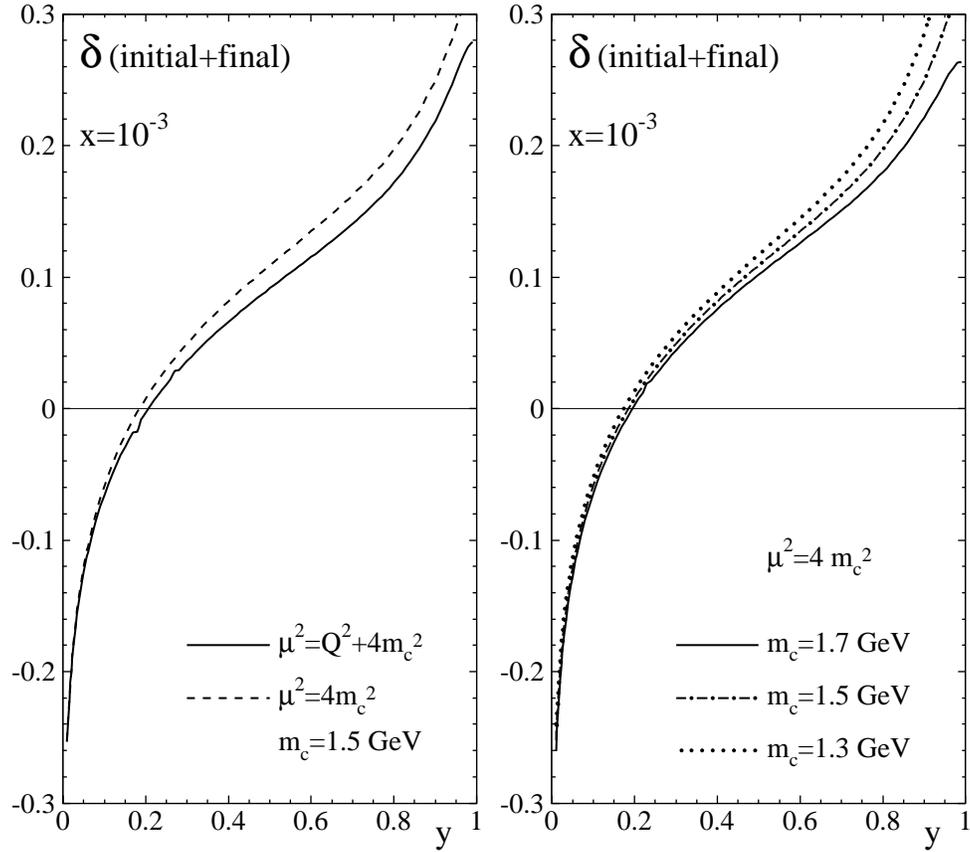,width=14cm}
\caption{\sf Left side: $\delta(x,y)$, for two different 
factorization scales 
$\mu^2=Q^2+4 m_c^2$ \protect\cite{vogt96} (full line) 
and $\mu^2=4 m_c^2$ \protect\cite{grs94} (dashed line) 
($m_c=1.5\ {\rm GeV}$).
Right side: $\delta(x,y)$, using the charm masses 
$m_c=1.3$, $1.5$, $1.7$~GeV 
($\mu^2=4m_c^2$).(PDF: GRV94(LO) \protect\cite{grv94}.)}
\label{compare}
\end{figure}

\begin{figure}[ht]
\centering
\vspace*{-0.2cm}
\epsfig{figure=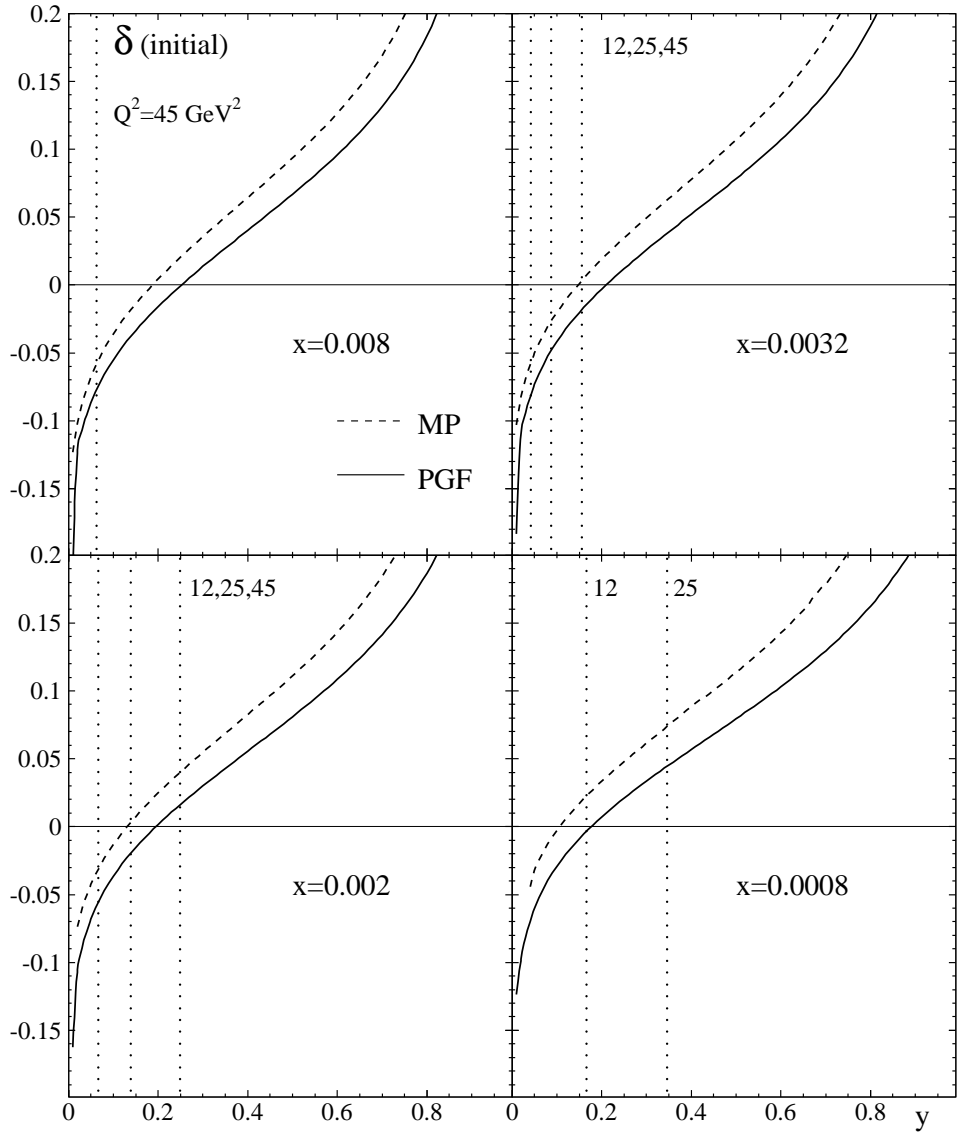,width=14cm}
\vspace*{-0.2cm}
\caption{\sf Comparison of massive 
(solid line, $\mu^2=Q^2+4m_c^2$, $m_c=1.5\ {\rm GeV}$) and 
massless (dashed line) radiative corrections $\delta$ (initial) 
for experimentally relevant $x$ \protect\cite{adloff96} using the 
parton distributions GRV92(LO) (MP) \protect\cite{grv92} and 
GRV94(LO) (PGF) \protect\cite{grv94}.
The dotted vertical lines indicate values of constant 
$Q^2=12$, $25$ or $45$ GeV$^2$.}
\label{mpcpgfgrvi}
\end{figure}


\begin{figure}[h]
\centering

\epsfig{figure=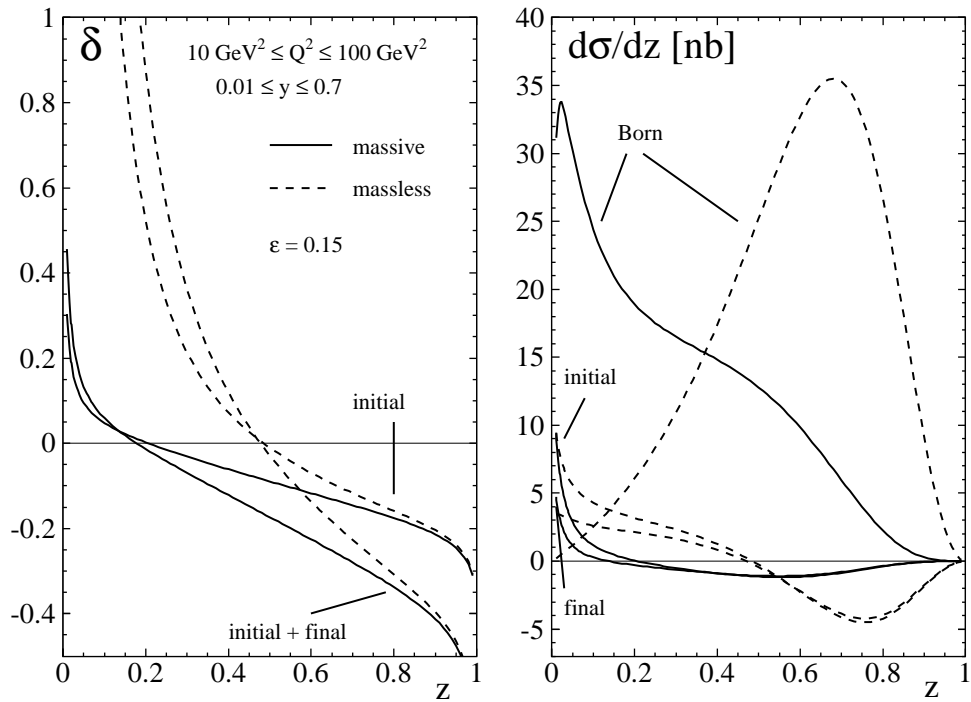,width=14cm}
\caption{\sf $z$-differential radiative corrections to massive 
($m_c=1.5\ {\rm GeV}$, $\mu^2=Q^2+4m_c^2$)
(full line) and massless (dashed) charm production, integrated 
over the kinematical range
$10\ {\rm GeV}^2\le Q^2\le 100\ {\rm GeV}^2,\ 0.01 \le y\le 0.7$, 
using the parton distributions GRV92(LO) (massless) 
\protect\cite{grv92} and GRV94(LO) (massive) \protect\cite{grv94}
and a Peterson et al.\ fragmentation function \protect\cite{peterson83}
with $\varepsilon=0.15$.}
\label{mpcpgffrag}
\end{figure}

\begin{figure}[ht]
\centering

\epsfig{figure=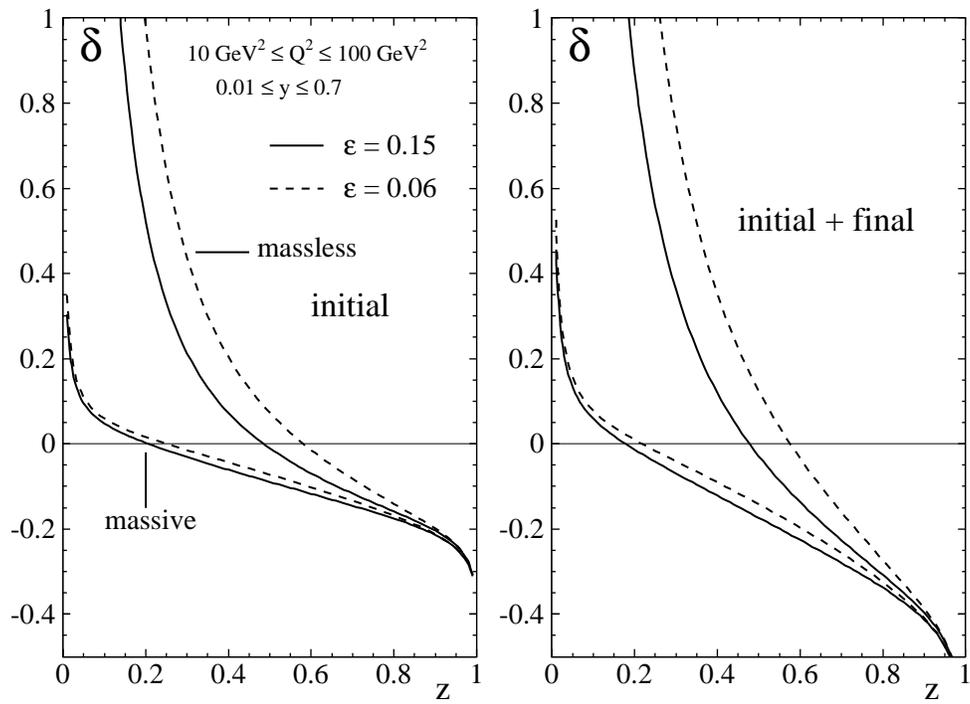,width=14cm}
\caption{\sf As fig.~\protect\ref{mpcpgffrag} (lhs)
with $\varepsilon=0.15$ (full line) and $\varepsilon=0.06$ 
(dashed line). 
The contribution due to initial state radiation is shown (lhs) 
as well as the sum of initial and final state radiation (rhs).}
\label{epscomp}
\end{figure}

\begin{figure}[ht]
\centering
\epsfig{figure=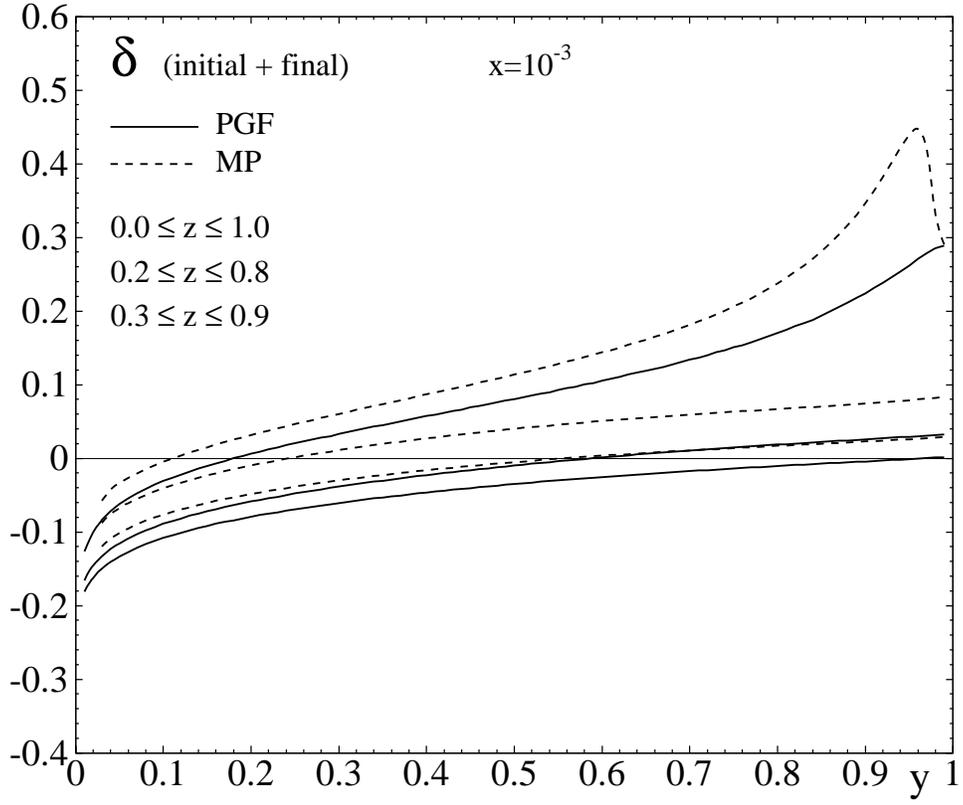,width=14cm}
\caption{\sf $\delta(x,y;z^{min},z^{max})$ (defined in the text) 
in dependence of $y$
for $x=10^{-3}$ and three choices of $z^{min}$, $z^{max}$:
From top to bottom
$0\le z\le 1$, $0.2 \le z \le 0.8$ and 
$0.3 \le z\le 0.9$.
The massive corrections have been calculated employing the
GRV94(LO) parton distributions \protect\cite{grv94} and the 
factorization scale $\mu^2=Q^2+4m_c^2$ with $m_c=1.5$ GeV.
For the massless corrections we have taken the
GRV92(LO) parton distributions \protect\cite{grv92}.
In all cases we employed a Peterson et al.\ fragmentation function 
\protect\cite{peterson83} with $\varepsilon=0.15$.}
\label{zcut1}
\end{figure}

\end{document}